# Observation of Temperature-Dependent Capture Cross-Section for Main Deep-Levels in β-Ga$_2$O$_3$

A.A. Vasilev [1], A.I. Kochkova [1], A.Y. Polyakov [1], A.A. Romanov [1], N.R. Matros [1], L.A. Alexanyan [1], I.V. Shchemerov [1] and S.J. Pearton [2]

1) National University of Science and Technology MISIS, Leninsky pr. 4, Moscow 119049, Russia
2) Department of Materials Science and Engineering, University of Florida, Gainesville, Florida 32611, USA

**Abstract:**

Direct observation of capture cross-section is challenging due to the need of extremely short filling pulses in the two-gate Deep-Level Transient Spectroscopy (DLTS). Simple estimation of cross-section can be done from DLTS and Admittance Spectroscopy (AS) data, but it is not feasible to distinguish temperature dependence of pre-exponential and exponential parts of the emission rate equation with sufficient precision conducting a single experiment. This paper presents experimental data of deep-levels in β-Ga$_2$O$_3$ that has been gathered by our group since 2017. Based on the gathered data we propose a derivation of apparent activation energy ($E_a^m$) and capture cross-section ($\sigma_n^m$) assuming temperature dependent capture via multiphonon emission model, which resulted in strong correlation between $E_a^m$ and $\sigma_n^m$ according to Meyer-Neldel rule, which allowed us to estimate low- and high-temperature capture coefficients $C_0$ and $C_1$ as well as capture barrier $E_b$. It also has been shown that without considering the temperature dependence of capture cross-section, the experimental values of $\sigma_n$ are overestimated by 1-3 orders of magnitude. A careful consideration of the data also allows to be more certain identifying deep-levels by their "fingerprints" ($E_a$ and $\sigma_n$) considering two additional parameters ($E_{MN}$ and $\sigma_{00}$) and to verify the density functional theory (DFT) computation of deep-level recombination properties.

1. **Introduction**

In recent years, Ga$_2$O has been actively investigated by numerous research groups due to its promising potential for power electronics and solar-blind UV detectors.[1,2] To advance the application of gallium oxide devices, detailed studies have been conducted on crystal growth,[1,3,4] epitaxial film growth,[1] intentional doping, and electrically active defects.[3] Our emphasis has been focused on the deep-level defects investigation by capacitance spectroscopy, mainly Deep-Level Transient Spectroscopy (DLTS)[5] and Admittance Spectroscopy (AS).[6,7] These two techniques are widely used for characterizing electrically active defects by evaluating their concentration ($N_t$), thermal activation energy ($E_a$), and electron capture cross section ($\sigma_n$).

These parameters are obtained from electron detrapping-kinetics characterized by the emission rate $e_n(T)$ dependence on temperature $T$, as given by:[5]

$$e_n(T) = \sigma_n \gamma T^2 \exp\left(-\frac{E_a}{kT}\right) \qquad (1)$$

where $\gamma = 4\sqrt{6} \cdot k^2 \pi^{3/2} h^{-3} m^*$, $m^*$ is an effective mass, $T$ is the temperature, $k$ the Boltzmann constant. Here $\sigma_n$ is assumed to be independent of temperature.

The kinetics of many thermally activated processes in chemistry and physics, including the emission rate of carriers from deep levels, are usually calculated from so called Arrhenius plots, which allow to extract the pre-exponential factor and activation energy as the temperature independent parameters for the sake of mathematical treatment simplification. This approach works well for simple estimation kinetics phenomena, however, if one is lacking knowledge of exact temperature dependence of pre-exponential factor and activation energy, it is not possible to reduce it to Arrhenius equation term by term and, therefore, the kinetics cannot be characterized with high precision. Furthermore, a capture cross-section and activation energy cannot be determined accurately.

1.1 **Meyer-Neldel rule in semiconductors**

In terms of semiconductor physics, this problem was faced by W. Meyer and H. Neldel in 1937 when they proposed empirical relationship given in equation (2) of the temperature dependence of pre-exponential factor for thermally activated conductivity.[8] This empirical law for the pre-exponential factor temperature dependence for thermally activated conductivity was presented and is now called Meyer-Neldel rule (MNR).

$$\sigma_0 = \sigma_{00} \exp\left(-\frac{E_a}{E_{MN}}\right) \qquad (2)$$

Various research groups confirmed this rule for carrier emission kinetics in A3B5 compounds,[9,10] ZnO and other semiconductors.[11,12] The most reasonable physical explanation of the MNR in terms of emission rate is based on considering the total change in Gibbs energy with entropy part as attributed to vibrational entropy $\Delta S_{\text{vib}}$. It has been concluded that vibrational entropy (estimated as rearrangement entropy of $n$ interacting phonons of $N$ total phonons in interaction volume) $\Delta S_{\text{vib}}/k_B = E_a/E_{MN} \cong n \ln(N/n)$ could not explain extremely small values of $\sigma_{00}$ which are ~$10^{-23}$ cm$^2$, and that without accumulation of a significant amount of data, it is not clear whether new ideas on this issue will be proposed.[11]

From the theoretical side, Alkauskas et. al. in 2014 developed[13] a theory on computing nonradiative capture cross-section for deep-level transitions occurring via multiphonon emission, which gives insights on defect recombinational properties at different temperatures and allows to identify defects from experimental data.

This current paper reveals new aspects of experimental deep-level parameters determination, provides results to verify such DFT defects computations, and demonstrates new data on capture coefficients for the main electron traps in β-Ga$_2$O$_3$.

## 2. Sampling

This analysis is based on DLTS and Admittance Spectroscopy data that has been gathered since 2017 for deep-levels in a wide range of β-Ga$_2$O$_3$ samples. The studied samples of β-Ga$_2$O$_3$ were cut from various types of wafers purchased from Tamura/Novel Crystals, Inc., Tokyo, Japan: β-Ga$_2$O$_3$ (-201) and (010) oriented edge defined film-fed grown (EFG) wafers doped with Sn, unintentionally doped EFG (-201) wafer, (010) oriented EFG wafers doped with Fe and (001) orientated unintentionally doped halide vapor phase epitaxy (HVPE) grown layers on bulk n$^+$-EFG substrates doped with Sn.[14] Different sets of treatments were employed to understand the presence and origin of electrically active deep levels in β-Ga$_2$O$_3$. A detailed description of the experiments and results along with the depiction of the deep-levels spectra could be found in our previous works.[15–21]

The sampling contains 1242 uncategorized data entities each corresponding to a single measurement of activation energy ($E_a^m$) and capture cross-section ($\sigma_n^m$) from Arrhenius plot in $\ln(e_n T^{-2})$ vs $1/T$ axes. The peak temperature ($T_{\text{peak}}$) is taken at the smallest window in the measurement and used only to improve quality of deep-level data clustering.

Up to now, a large number of groups have already done a significant work in the field of theoretical and experimental identification of electrically active defects in gallium oxide. Let us briefly describe the results of deep-levels characterization that are taken as reference data in this analysis.

Center E1 (with $E_a$ found in range (0,45÷0,65) eV, and $\sigma_n$ found in range (0.3÷7)·10$^{-13}$ cm$^2$) has been introduced by H-plasma treatment,[19] proton irradiation,[18,21] and ampoule annealing in H$_2$.[15,22] It has been observed that the E1 is a donor, and according to the theoretical models, a possible configuration is the complex of H with shallow donors Si or Sn.[15]

The center E2 ($E_a$=(0,74÷0,82) eV, $\sigma_n$=(0.6÷23)·10$^{-15}$ cm²) is often detected in EFG, HVPE grown samples and assigned to Fe acceptors.[23–32] The E2* centers ($E_a$=(0,75÷0,78) eV, $\sigma_n$=(2÷7)·10$^{-14}$ cm²) typically have been observed after radiation or implantation and demonstrates a linear increase with irradiation exposure.[21] This implies the possible origin of E2 , which is a complex of intrinsic point defects of gallium and oxygen vacancies.[32]

For the E3 ($E_a$=1.05 eV, $\sigma_n$=4.1·10$^{-13}$ cm²) level detected in unintentionally doped EFG-grown β-Ga$_2$O$_3$, it has been suggested that the possible nature of the center is a deep donor related to Ti.[33] However, a defect with a similar $E_a$ and $\sigma_n$ tends to increase in concentration after irradiation with high-energy particles (neutrons and protons) and Ar plasma treatment.[21] So, the issue with these two interpretations could be the same as for E2 and E2* at early stages of β-Ga$_2$O$_3$ research.[32]

The E8 ($E_a$=0,28 eV, $\sigma_n$=6·10$^{-18}$ cm²) center is an intrinsic point defect or complex detected after irradiations and treatment with Ar and H plasma.[21]

## 2.1 Deep-level clustering

Data clustering was performed using a Gaussian Mixture model with Variational inference algorithm.[34] This method assumes all data can be represented by finite mixture of Gaussian distributions with unknown parameters which are determined from a variational lower boundary. The above procedure reduced the total data entities from 1242 to 1033, excluding dropouts, and produced 6 clusters assigned as main deep-levels E2*, E4, E8, E1, E2 and E3. Results of data clustering can be seen in the pairwise plot in Figure 1, where normalized distributions of $\sigma_n^m$, $E_a^m$ and $T_{\text{peak}}$ for each trap are presented on the main diagonal plots and pairs of parameters plotted pairwise on off-diagonal plots.

Variances and mean values of $\sigma_n^m$, $E_a^m$ and $T_{peak}$ for each deep-level can be determined from the data presented in Figures 1a, 1e and 1i. The vertical alignment of clusters on Figures 1c, and 1f demonstrates no correlation of $\sigma_n^m$ and $E_a^m$ with $T_{peak}$ (since $\sigma_n^m$ and $E_a^m$ are computed from Arrhenius plots in temperature ranges higher than $T_{peak}$ variance) but there is strong correlation of $\sigma_n^m$ with $E_a^m$ on Figure 1b. This relationship will be used further for equation (2) fit and results analysis.

## 3. Derivations

Typically, the most straightforward and accurate way to determine capture cross-section is through direct observation of capture kinetics, but this method necessitates the usage of extremely short pulsing times even for materials with low doping levels in the two-gate DLTS[35] ($\tau_p^{-1} = C_n \cdot n \approx 10^{-9} \cdot 10^{15} = 1$ MHz). More simple ways are to calculate the capture cross-section from standard DLTS and AS approaches,[5–7] but these techniques are not suitable for precise capture cross-section measurements, especially with the assumption of its strong temperature dependance. Emission rates can be measured by implementing long and short windows for DLTS to compute the low- and high-temperature cross-section from Arrhenius plots, but in this case, it is not feasible to separate the temperature dependence of pre-exponential and exponential parts with sufficient precision conducting a single experiment.

The following sections will reveal a treatment of this issue which allows to extract activation energy ($E_a$), low- and high-temperature capture coefficients ($C_0$, $C_1$), and capture barrier ($E_b$) of the deep-level through detailed DLTS and AS data analysis.

### 3.1 Vibrational entropy

The entropy term, as mentioned in the introduction, was proposed to explain the observed correlation of $E_a$ and $\sigma_n$ for other semiconductor materials. In addition, its appearance in Gibbs free energy ($-\Delta G/kT = -\Delta H/kT + \Delta S/k$) fits well with temperature independent behavior described by equation (2), but the vibrational entropy is relatively small $\Delta S_{\text{vib}}/k = \pm(1.5 \div 3)$[36,37] when atomic rearrangements around defect are neglected.

## 3.2 Carrier capture by multiphonon emission

The attempt to consider coupling with local mode $Q$ can be done with the given temperature dependence of carrier capture via multiphonon emission. The capture coefficient can be written in its low-temperature form (3),[38] with $p$ and $S$ defined from configurational diagram (Fig. 2)

$$C_n \propto (\bar{n}+1)^p \frac{S^p}{p!} \exp\left(-2S\left(\bar{n}+\frac{1}{2}\right)\right) \qquad (3)$$

The expression (3) can be approximated at low and high temperatures with the equation (4):[13,38]

$$C_n(T) \approx C_0 + C_1 \exp\left(-\frac{E_b}{kT}\right) \qquad (4)$$

And then capture cross-section by definition will be:

$$\sigma_n(T) = C_n(T)/\langle v_{th}\rangle \qquad (5)$$

where $\langle v_{th}\rangle = \sqrt{3kT/m^*}$ – average thermal velocity. Thus, the temperature dependence of emission rates for carriers trapped at deep-level ($E_a = E_k(0) - E_d(Q_d)$) will be:

$$e_n = \sigma_n(T) \cdot \gamma T^2 \exp\left(-\frac{E_a}{kT}\right) \qquad (6)$$

Since the emission rate data from DLTS/AS experiment is treated with Arrhenius plot, we need to analytically keep temperature dependence of (6) until we will be ready to understand nature of observed $E_a^m$ and $\sigma_n^m$ correlation (Fig. 1b) and then simplify resulted expression in order to estimate coefficients of equation (4). Arrhenius plot (Fig. 3a) in $\ln(e_n T^{-2})$ vs $1/T$ axes after collecting results from equations (4), (5) and (6) will be:

$$\ln(e_n T^{-2}) = -\ln\langle v_{th}\rangle + \ln\gamma + \ln\left(C_0 + C_1 \exp\left(-\frac{E_b}{kT}\right)\right) - \frac{E_a}{kT} \qquad (7)$$

Therefore, apparent activation energy is computed from the slope of a tangent line at temperature $T$ in $\ln(e_n T^{-2})$ vs $1/T$ axes. This slope as a function of temperature will be:

$$\text{slope}(T) = \frac{d}{d(1/T)}\ln(e_n T^{-2}) = \frac{1}{2}T - \frac{E_b/k \cdot C_1}{C_0 e^{\frac{E_b}{kT}} + C_1} - \frac{E_a}{k} \qquad (8)$$

And corresponding intercept of a tangent line $(\text{intercept}(x) = f(x) - x \cdot f'(x))$ at temperature $T$, from which apparent capture cross-section is computed in $\ln(e_n T^{-2})$ vs $1/T$ axes will be:

$$\text{intercept}(T) = -\ln\langle v_{th}\rangle + \ln\gamma + \ln\left(C_0 + C_1\exp\left(-\frac{E_b}{kT}\right)\right) - \frac{E_a}{kT} - \frac{1}{T}\cdot\text{slope}(T) \qquad (9)$$

Eventually, measured $E_a^m$ and $\sigma_n^m$ from slope and intercept at some temperature $T$ will be:

$$E_a^m = -\text{slope}(T)\cdot k, \qquad (10)$$

$$\sigma_n^m = \frac{\exp(\text{intercept}(T))}{\gamma}. \qquad (11)$$

It can be clearly seen from (8) and (10) that $E_a^m$ appears to be step-like function (Fig. 3b) of $T$, with low- and high-temperature plateaus that are roughly $E_a$ and $E_a + E_b$, with shift due to the presence of $-kT/2$ in $E_a^m$. More precise values for lower and upper boundaries are $(E_a - \Delta)$ and $(E_a + E_b - \Delta)$ where shift value $\Delta$ is $\Delta = -E_b(\ln(C_1/C_0) - 2)/(2\ln^2(C_1/C_0))$. The same temperature behavior of $\ln(\sigma_n^m)$ is presented on Fig. 3c with the lower and upper boundaries at the same temperatures as $E_a^m$.

In order to reproduce results from Fig 1b, the derivative of $\ln(\sigma_n^m)$ with respect to $E_a^m$ will be:

$$\frac{d(\ln(\sigma_n^m))}{dE_a^m} = \frac{(\partial \ln(\sigma_n^m)/\partial T) \cdot dT}{(\partial E_a^m/\partial T) \cdot dT} = \frac{1}{kT} \tag{12}$$

Summing up the intermediate results, (10) and (11) implies that $E_a^m$ and $\ln(\sigma_n^m)$ correlates through the temperature at which DLTS/AS can be performed, but experimentally we are limited in the analysis for the same reason elaborated in section 2.1 – data points from DLTS/AS data taken in broad temperature interval (around $\Delta T \approx 50K$ for E2 trap) when noticeable changes in $E_a^m$ or $\ln(\sigma_n^m)$ are happening in narrow temperature range. This narrow temperature range appears form data linearity in Fig 1b implying by (12) ($1/kT$ should be constant or vary slow to give linearity in Fig. 1b).

The gathered data for E2 trap (Fig. 3b,c) shows that changes in $\ln(\sigma_n^m)$ occur in a narrow temperature region near some temperature $T_m$ so almost linear data in $\ln(\sigma_n^m)$ vs $E_a^m$ axes will be observed. To estimate $T_m$ it can be assumed that $E_a^m(T_m)$ and $\sigma_n^m(T_m)$ corresponds to the middle of the step of (10) and (11) leading to same result for $T_m$ (Fig. 3b,c):

$$E_a^m(T_m) = E_a + \frac{E_b}{2} \Rightarrow T_m = \frac{E_b}{k \ln(C_1/C_0)} \tag{13}$$

$$\frac{C_1}{C_0} \exp\left(-\frac{E_b}{kT_m}\right) = 1 \Rightarrow T_m = \frac{E_b}{k \ln(C_1/C_0)} \tag{14}$$

Since $E_a^m$ and $\ln(\sigma_n^m)$ data shows no correlation to $T_{\text{peak}}$ due to experiment limitations, and the biggest changes in measured parameters appears around $T_m$, we can expand $E_a^m(T)$ and $\ln \sigma_n^m(T)$ near $T_m$, excluding temperature from equations and establishing functional dependence in form $\ln(\sigma_n^m) = f(E_a^m)$.

So, expanding $E_a^m$ at $T_m$ up to the linear term:

$$E_a^m(T)\big|_{T_m} \approx E_a^m(T_m) + \frac{dE_a^m}{dT}\bigg|_{T_m}(T-T_m) + O(T^2)$$
$$= E_a + E_b\left(\frac{1}{2} - \frac{\ln(C_1/C_0)}{4}\right) + kT\left(\frac{\ln^2(C_1/C_0)}{4} - \frac{1}{2}\right) \quad (15)$$

And for $\ln(\sigma_n^m)$ at $T_m$:

$$\ln \sigma_n^m(T)\big|_{T_m} \approx \ln \sigma_n^m(T_m) + \frac{d\ln \sigma_n^m}{dT}\bigg|_{T_m}(T-T_m) + O(T^2)$$
$$= \ln(2C_0) - \frac{1}{4}\left(\ln^2\left(\frac{C_1}{C_0}\right) - 2\ln\left(\frac{C_1}{C_0}\right)\right) - \frac{1}{2}\ln\left(\frac{3E_b}{m^*\ln(C_1/C_0)}\right) \quad (16)$$
$$+ \frac{kT}{E_b}\left(\frac{\ln^3(C_1/C_0)}{4} - \frac{\ln(C_1/C_0)}{2}\right)$$

Expressing and equating $T$ from (15) and (16) brings us to $\ln(\sigma_n^m) = f(E_a^m)$ near $T_m$, which allows to extract deep-level parameters from fitting the resultant expression for red line in Fig. 3:

$$f(x) = \ln(2C_0) - \frac{E_a}{E_b}\ln\left(\frac{C_1}{C_0}\right) - \frac{1}{2}\ln\left(\frac{3E_b}{m^*\ln(C_1/C_0)}\right) + x\frac{\ln(C_1/C_0)}{E_b} \quad (17)$$

From $f(x=0)$ condition low-temperature capture coefficient $C_0$ can be estimated with experimentally known $\sigma_{00}$:

$$\sigma_{00} = \frac{2C_0}{\sqrt{\frac{3kT_m}{m^*}}}\exp\left(-\frac{E_a}{kT_m}\right) \quad (18)$$

### 3.3 Model fitting

To find all 4 ($E_a$, $E_b$, $C_0$, $C_1$) model parameters from experimental data it is needed to impose two more restrictions besides (13) and (18).

Assuming that enough data from many different time windows has been gathered and, therefore, the mean values of this data are represented by $E_a^m(T_m)$ and $\ln \sigma_n^m(T_m)$, the complete system of equations can we written:

$$\begin{cases} E_a + \dfrac{E_b}{2} \approx \text{mean}(E_a^m), & \text{(i)} \\[6pt] \log_{10}\left(\dfrac{2\sqrt{C_1 C_0}}{\sqrt{3kT_m/m^*}} \exp(-1/2)\right) \approx \text{mean}\left(\log_{10}(\sigma_n^m)\right), & \text{(ii)} \\[6pt] \dfrac{E_b}{k \ln(C_1/C_0)} = T_m, & \text{(iii)} \\[6pt] \log_{10}\left(\dfrac{2C_0}{\sqrt{3kT_m/m^*}} \exp\left(-\dfrac{E_a}{kT_m}\right)\right) = \log_{10}(\sigma_{00}). & \text{(iv)} \end{cases} \quad (19)$$

This system can't be solved analytically, so it might be suggested to use the numerical approach of finding the solution by minimizing the following function:

$$F = \sum_{i=1}^{4} f_i - f_i^{\text{RHS}} \qquad (20)$$

And corresponding solution will be:

$$E_a, E_b, C_0, C_1: \min_{E_a, E_b, C_0, C_1} \{\|F\|_2^2\} \leq 10^{-5} \qquad (21)$$

### 4. Results

Computed $E_{MN} = kT_m$ and $\sigma_{00}$ for deep-levels E2, E2*, E1, E8 and E3 (E4 is omitted here due to the lack of extensive collected data) in β-Ga$_2$O$_3$ are presented in Fig. 4. This data is then used for solving Eq. (21) to provide the right-hand side for equation (iii) and (iv) of (19).

The distributions of gathered $E_a^m$ and $\sigma_n^m$ are represented in box-plots on Fig. 5. Boxes on Figure 5 represent $\text{mean}(E_a^m)$ and $\text{mean}(\log_{10}(\sigma_n^m))$ at temperature $T_m$ and used as right-hand side of equations (i) and (ii) of (19). Solutions of (21) for each trap are presented in Table 1 and plotted in Figure 6.

In comparison with Fig. 5, activation energies and capture cross-sections obtained in experiments are overestimated due to interplay of the temperature dependence of capture cross-section and the thermal emission terms, and, in this case, the previously measured values of capture cross-section (around $T_{\text{peak}}$) are 1-3 orders of magnitude higher than obtained with suggested model.

Approaching the same problem from theoretical side of the issue, similar results were obtained by Wickramaratne et al. [39] In their paper, computed temperature dependent capture cross-section via multiphonon emission mechanism was inserted into DLTS formalism, and it appeared to shift apparent activation energy obtained from Arrhenius plot to higher values at higher temperatures. Nevertheless, the study[39] was not supported by any experimental data, unlike the present paper, which is possibly the first experimental observation on this matter.

## 5. Summary and Conclusions

In this paper we have considered that the carrier emission rate from main deep-levels in β-Ga$_2$O$_3$ follow the MN-rule and it has been shown that the capture cross section in multiphonon emission model explains the observed $E_a^m$ and $\sigma_n^m$ shift.

We have applied the theory to the main deep-level centers in β-Ga$_2$O$_3$ and accurately calculated all parameters (Table 1), including activation energy $E_a$, barrier height for carrier capture $E_b$, as well as low- and high-temperature capture coefficients $C_0$ and $C_1$.

This suggests using of $E_{MN}$ and $\sigma_{00}$ as two additional parameters to identify defects, employing of $C_0$, $C_1$ and $E_b$ to estimate capture coefficient with more detailed and advanced approach, and to verify DFT results on recombinational properties of deep levels.

**Acknowledgments**

The research at NUST MISIS was funded by Ministry of Science and Higher Education of the Russian Federation grant number 075-15-2022-1113. The work at UF was performed as part of Interaction of Ionizing Radiation with Matter University Research Alliance (IIRM-URA), sponsored by the Department of the Defence, Defence Threat Reduction Agency under award HDTRA1-20-2-0002. The content of the information does not necessarily reflect the position or the policy of the federal government, and no official endorsement should be inferred.

**Data availability statement**

The data that support the findings of this study are available from the A.A. Vasilev, upon reasonable request.

**Table 1.** Results of model fitting for main deep-levels in β-Ga$_2$O$_3$

| Trap | $E_a$ (eV) | $E_b$ (eV) | $C_0$ (cm$^3$/s) | $C_1$ (cm$^3$/s) |
|------|-----------|-----------|------------------|------------------|
| E2   | 0.68      | 0.21      | 7.0×10$^{-10}$   | 1.4×10$^{-6}$    |
| E2*  | 0.61      | 0.21      | 1.3×10$^{-9}$    | 2.6×10$^{-6}$    |
| E1   | 0.52      | 0.17      | 1.3×10$^{-8}$    | 3.1×10$^{-5}$    |
| E8   | 0.23      | 0.12      | 2.0×10$^{-11}$   | 5.3×10$^{-8}$    |
| E3   | 0.89      | 0.19      | 2.0×10$^{-8}$    | 2.0×10$^{-5}$    |

**Figure Captions**

Figure 1. Pairwise plot of clustered data for main deep-levels in β-Ga$_2$O$_3$. (a,e,i) – distributions of experimental parameters for trap clusters. (b,c,f) and (d,g,h) – represents the same data and shows correlation within measured parameters. These scatter plots demonstrate no correlation of $\sigma_n^m$ or $E_a^m$ with $T_{\text{peak}}$, which variance is determined with technique limitations. However, $\sigma_n^m$ strongly depends on $E_a^m$ and this phenomenon will be studied more thoroughly in sections below. (j) – one of the DLTS measurements from the gathered data, which provides 3 of 1242 data entities represented with Arrhenius plot (k) and attributed to E1, E2 and E4 levels.

Figure 2. Configurational diagram. E$_d$ – energy of electron on a defect with Q = Q$_d$ (non-equal to 0 with presence of electron-lattice interaction) and E$_k$ – exited state which is delocalized and since then Q = 0 (small electron-lattice coupling to local mode Q).

Figure 3. Arrhenius plot and extracted parameters as a function of temperature. Fully analytical model (gradient line) with the accounting of temperature dependence of capture cross-section as eq. (5) and simplified model eq. (15) (red line) of temperature dependence of measured $E_a^m$ and $\sigma_n^m$. (a) – Fully analytical Arrhenius plot showing different slope values at low and high temperature, (b) and (c) – $E_a^m$ and $\sigma_n^m$ calculated from Arrhenius plot showing step-like function and (d) – parametric plot of $\sigma_n^m$ and $E_a^m$ with full and simplified models based on our E2 data (violet crosses) and others groups data (blue symbols).

Figure 4. Collected data on main deep-levels in β-Ga$_2$O$_3$ (magnified plot from Fig 1b) and fit with parameters from Table 1.

Figure 5. $E_a^m$ and $\sigma_n^m$ distributions of main deep-levels in β-Ga$_2$O$_3$, based on collected data.

Figure 6. Fitted model of main deep-levels in β-Ga$_2$O$_3$ (data from Table 1).

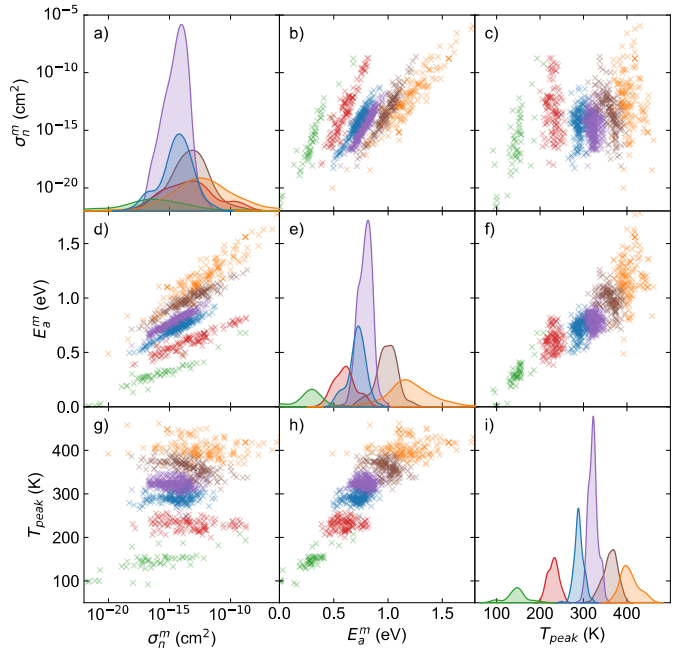
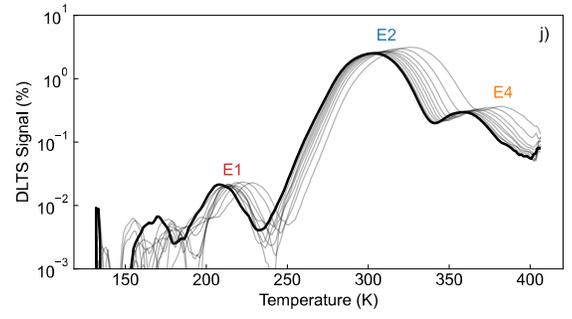
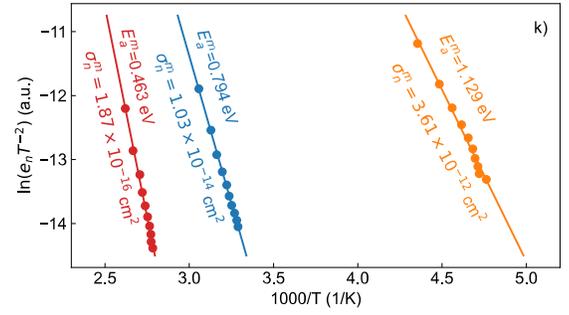

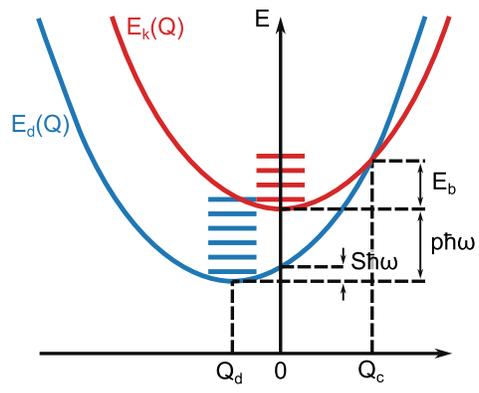

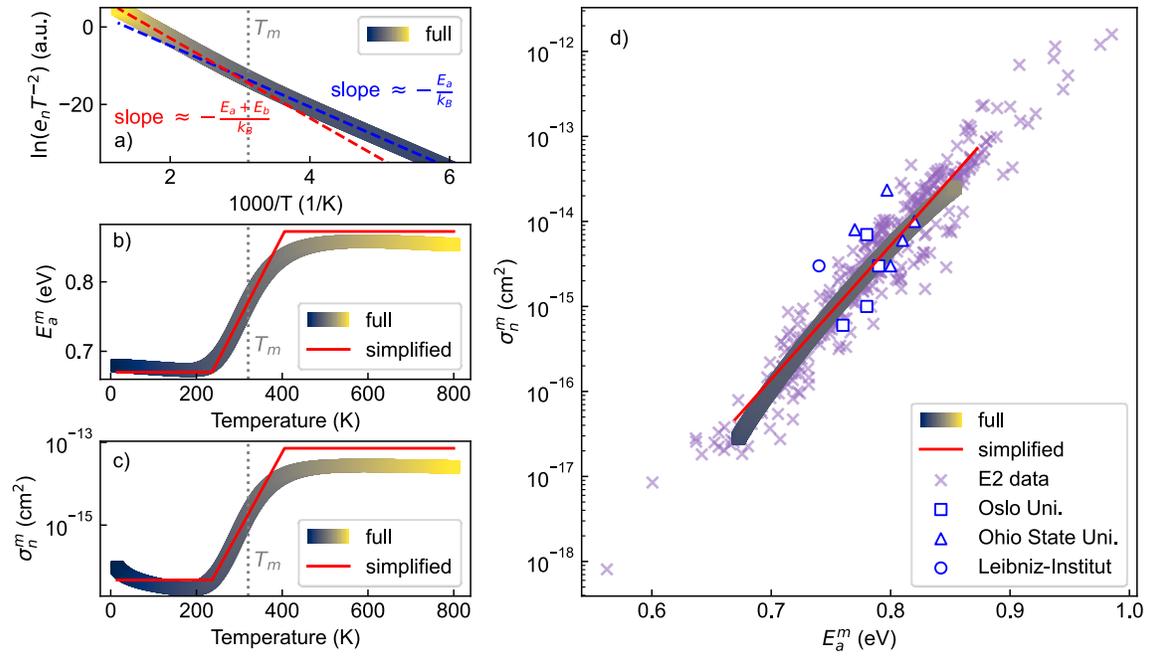

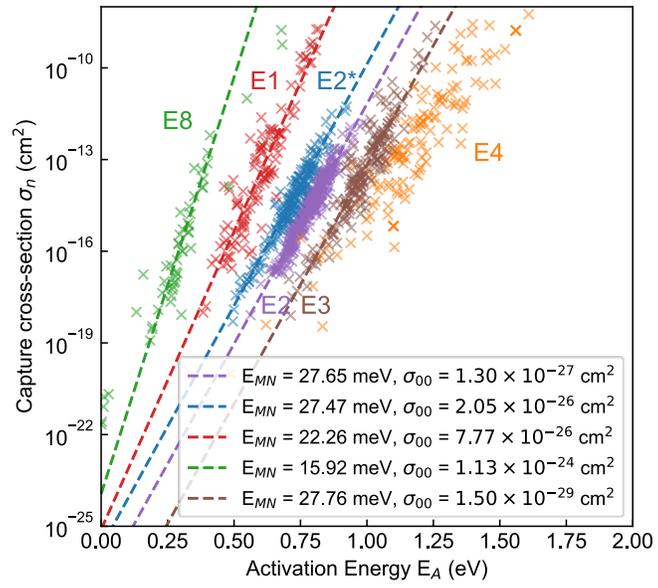

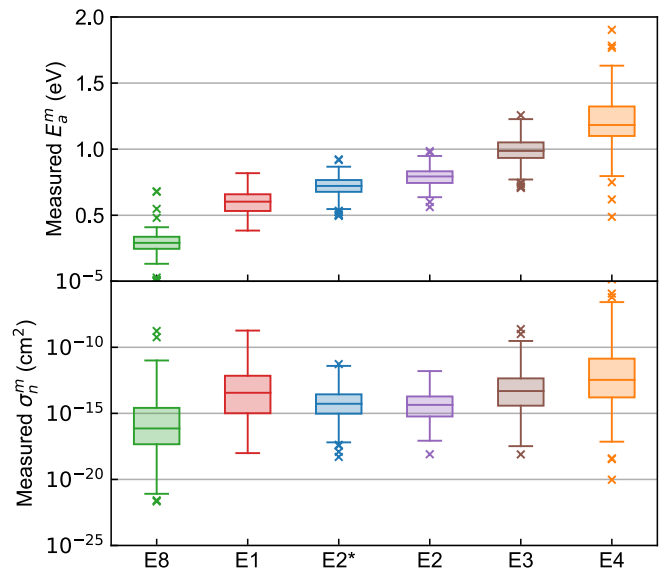

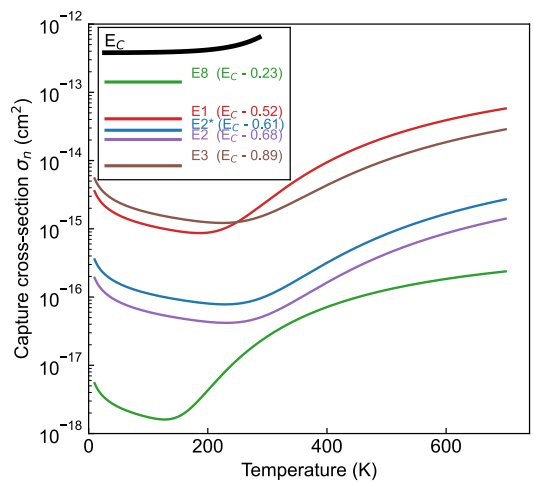